\documentclass[sigconf,authorversion,nonacm,screen]{acmart}
\usepackage{amsthm}
\usepackage{amsmath}
\usepackage{nicefrac}

\usepackage{graphicx}
\usepackage[caption=false]{subfig}

\usepackage{url}
\usepackage{caption}

\usepackage{algorithm}
\usepackage{algpseudocode}

\usepackage{tikz}
\tikzset{mynode/.style = {draw,circle,inner sep=0pt, minimum size=0.4cm}
         }

\makeatletter
\def\plist@algorithm{Alg.\space}
\makeatother

\usepackage{graphicx}
\usepackage{tikz}

\usepackage{multirow}

\newcommand{\att}{\mathbf{A}}

\AtBeginDocument{%
  \providecommand\BibTeX{{%
    \normalfont B\kern-0.5em{\scshape i\kern-0.25em b}\kern-0.8em\TeX}}}

\setcopyright{rightsretained}

\begin{document}

\copyrightyear{2021} 
\acmYear{2021} 
\acmConference[ARES 2021]{The 16th International Conference on Availability, Reliability and Security}{August 17--20, 2021}{Vienna, Austria}
\acmBooktitle{The 16th International Conference on Availability, Reliability and Security (ARES 2021), August 17--20, 2021, Vienna, Austria}\acmDOI{10.1145/3465481.3465761}
\acmISBN{978-1-4503-9051-4/21/08}
\title{How Lightning's Routing Diminishes its Anonymity}

\author{Satwik Prabhu Kumble}
\affiliation{%
  \institution{TU Delft}
  \country{Netherlands}
}

\author{Dick Epema}
\affiliation{%
  \institution{TU Delft}
  \country{Netherlands}}

\author{Stefanie Roos}
\affiliation{%
  \institution{TU Delft}
  \country{Netherlands}
}

\renewcommand{\shortauthors}{Prabhu Kumble, et al.}

\begin{abstract}
  Lightning, the prevailing solution to Bitcoin’s scalability issue, uses onion routing to hide senders and recipients of payments. Yet, the path between the sender and the recipient along which payments are routed is selected such that it is short, cost efficient, and fast. The low degree of randomness in the path selection entails that anonymity sets are small. However, quantifying the anonymity provided by Lightning is challenging due to the existence of multiple implementations that differ with regard to the path selection algorithm and exist in parallel within the network.
  \\
  In this paper, we propose a general method allowing a local internal attacker to determine sender and recipient anonymity sets. Based on an in-depth code review of three Lightning implementations, we analyze how an adversary can predict the sender and the recipient of a multi-hop transaction. Our simulations indicate that only one adversarial node on a payment path uniquely identifies at least one of sender and recipient for around 70\% of the transactions observed by the adversary. Moreover, multiple colluding attackers can almost always identify sender and receiver uniquely.
\end{abstract}





\maketitle

\section{Introduction}
\label{sec:intro}

Payment channel networks like Lightning~\cite{lightning} promise high scalability, low latency, and low fees. In this manner, they constitute one of the most auspicious approaches to overcoming the limitations of Proof-of-Work blockchains such as Bitcoin~\cite{bitcoin} and Ethereum~\cite{wood2014ethereum}.
In addition to the performance advantages, payment channel networks supposedly also improve user privacy as transactions are executed locally and not recorded in the blockchain. Yet, the claim of improved privacy comes with a lack of in-depth evaluation of novel privacy risks~\cite{gudgeon2019sok}.

Two blockchain users open a payment channel by locking collateral on the blockchain. 
Afterwards, they can locally conduct transactions with each other arbitrarily often as long as the transaction values do not exceed the locked collateral. 
These transactions change the balance of the channel, i.e., the distribution of the collateral between the two parties. 
Each of the two parties can decide to close the channel unilaterally and then both parties obtain the funds corresponding to the current balance of the channel. 
Hence, direct interaction with the blockchain, which corresponds to global dissemination and consequentially often low privacy~\cite{meiklejohn2013fistful,biryukov2014deanonymisation}, is only necessary for opening and closing channels as well as for resolving disputes about the channel balance~\cite{gudgeon2019sok}. Thus, direct transactions in payment channels indeed preserve privacy. 

\begin{figure*}
    \centering
    \vspace{-2.5em}
    \subfloat[Shortest Path Routing]{\includegraphics[scale=0.5]{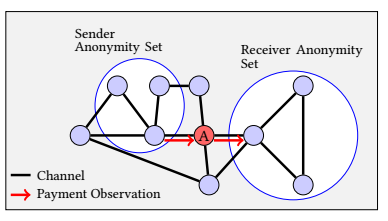}\label{fig:ex1}}\hfill 
    \subfloat[Shortest Path Routing with Timelocks]{\includegraphics[scale=0.5]{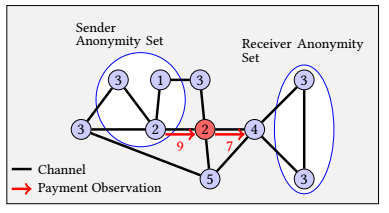}\label{fig:ex2}}\hfill
    \vspace{-1.5em}
    \caption{Example of inferring the sender and recipient anonymity sets for routes corresponding to shortest paths. The attacker observes the payment forwarded according to the red arrows. As the transaction follows a shortest path, the attacker can determine the potential senders and recipients as all nodes for which a shortest path contains the two edges on which the attacker observes the payment. This attack is possible for any source routing protocol using shortest paths or  similar selection strategies. However, Lightning provides additional information in the form of timelock values chosen by nodes, as indicated by the numbers in b). The attacker node knows the sum of the timelocks (TTL=9 before reaching the attacker and TTL'=7 after being forwarded by the attacker) of the subsequent nodes on the path and can hence reduce the sender and recipient anonymity sets accordingly. The recipient node has to be a node such that all timelocks along a path from the attacker to the recipient add up to 9 (including the attacker's timelock)}
    \label{fig:ex}
    \vspace{-2em}
\end{figure*}

However, rather than opening a channel for every transaction partner to conduct a direct transaction, a source or sender can also route a payment via multiple intermediaries to a destination or recipient. In such a payment, intermediaries are involved in the payment and achieving privacy properties such as anonymity for the sender and the recipient becomes more challenging. 
Bitcoin's Lightning~\cite{lightning} and Ethereum's Raiden~\cite{raiden} both apply onion encryption to hide the sender and the recipient from intermediaries. They both apply source routing. The sender determines the payment path and obtains the public keys of all the nodes along the path. The sender then uses layered encryption such that each intermediary only learns the identity of its predecessor and successor on the path\footnote{\url{github.com/lightningnetwork/lightning-rfc/blob/master/04-onion-routing.md}}. 
Despite Lightning's onion encryption being called Tor-style by its developers~\footnote{\url{https://coinjournal.net/bitcoin-developers-explain-tor-style-onion-routing/}}, the two protocols are extremely different with regard to their privacy guarantees. Whereas Tor selects random routers, Lightning chooses one of the least costly paths in terms of a cost metric related to fees, path length, and other globally known path properties. Intuitively, using such a strategic\footnote{we use the term strategic rather than deterministic as there is a low degree of randomness in some of the applied routing strategies} routing protocol results in much smaller anonymity sets than random path selection.

It has been shown that the anonymity of the sender and recipient can be compromised if the nodes right after the sender and right before the recipient in a payment path are controlled by an adversary ~\cite{tikhomorov_snp}. However, the evaluation was limited in the sense that only paths of up to length 5 was allowed between any pair of nodes. Other works evaluating the anonymity in Bitcoin's Lightning network studied the effectiveness of inferring the receiver through timing attacks~\cite{nisslmueller2020toward}, linkability of payments at different locations in the network~\cite{malavoltamulti}, and an attack that determines whether the predecessor is indeed the source~\cite{kappos2020empirical}. The only attack that explicitly makes use of Lightning's predictable routing targets availability rather than anonymity~\cite{tochner2019hijacking}.

In this work, we design an attack that predicts the sender and recipient of a transaction in the Lightning Network by leveraging our knowledge of the routing algorithms. It is important to note that there are multiple Lightning clients with different routing protocols. All of these routing protocols exist in parallel in the network. Our attack is designed such that it does not only consider all the existing clients, it can furthermore easily be extended to include novel client implementations.

In a nutshell, our attack places one or multiple attackers in the Lightning Network who observe transactions going through them. The key concept required for the attack is the timelock. If there is a dispute about a channel's state, nodes have to raise a dispute within a certain timelock. Nodes can individually set that timelock in terms of number of Bitcoin blocks added to the blockchain. During a multi-hop payment, the sender computes the sum of all timelocks of channels on the path excluding itself. It then uses this sum as the timelock for the payment to its successor on the path. When an intermediary forwards a payment, it decreases the total timelock value by its local timelock and then forms a commitment to make a payment to its successor during this reduced time. 
In this manner, an intermediary node with dispute timelock $x$ still has time $x$ to raise a dispute even if the remaining hop-by-hop payments on the path take maximal time~\cite{lightning}. 
For intermediaries to be able to ensure that they indeed have sufficient time to raise disputes, the sum of subsequent timelocks is hence included with the transaction they forward. 
In addition, the timelocks of individual channels and the network topology are publicly known.
The attacker combines this public information with the timelocks associated with the transaction to infer potential recipients. For each such recipient, the attacker then finds all the possible senders that could choose a path through the attacker to reach the recipient.

Figure~\ref{fig:ex} presents an example on how the available information can be used to narrow down potential senders and recipients. In contrast to the multifaceted cost functions utilized by Lightning's clients to select paths, we illustrate the principle by choosing shortest paths. As for source routing for anonymous communication on the network layer~\cite{chen2015hornet}, an attacker can determine  potential senders and recipients as those nodes that have shortest paths containing the two edges adjacent to the attacker over which the payment traverses. However, the timelock information in Lightning allows to further narrow down the anonymity sets. Thus, our attack on Lightning goes beyond attacks on source routing in other contexts as payment channel networks reveal very application-specific information. In addition to timelocks, fees and payment amounts assist in decreasing the anonymity set size.

Our simulations using a real-world Lightning snapshot indicate that the size of the anonymity sets determined by the attacker is indeed low. We chose 21 adversaries using various criteria. Together, these adversaries observed slightly more than 50\% of the transactions. In 70\% of cases a node observed a transaction, they were able to uniquely identify the sender or recipient.
Additionally, both the sender and the recipient could be uniquely identified in about 8\% of the attacks. For transactions passing more than one adversary, the recipient was always uniquely identified when all adversaries share their respective anonymity sets and sender was uniquely identified in excess of 40\% such cases.

In summary, we show that anonymity in multi-hop Lightning payments is low, as previously suspected but never quantified. However, as we discuss for multiple protection mechanisms, defenses come at a high cost in terms of effectiveness and efficiency.

\section{Lightning}
\label{sec:background}

We introduce Lightning, Bitcoin's payment channel network. Lightning was launched in 2016 and has nearly 10000 active nodes and 40,000 channels, as of March 25 2021\footnote{\url{https://1ml.com/statistics}}. There are three active clients: Lightning Labs' $LND$\footnote{\url{https://lightning.engineering/}}, ACINQ's $Eclair$\footnote{\url{https://acinq.co/}}, and Blockstream's $c-Lightning$\footnote{\url{https://blockstream.com/lightning/}}. Clients differ slightly with regard to their implementation of routing, though they all share the same key idea of routing a payment along a short and cheap path. 

While our evaluation is specifically targeted towards Lightning, the key idea of the attack is equally applicable to other payment channel networks that use source routing with a strategic selection of routes. Indeed, it is possible to extend our attack to include new clients with different routing strategies as long as they follow the general Lightning specifications with regard to the choice of fees and timeouts. 

\subsection{Lightning Payment Channels}

We model the Lightning network using graph-theoretical notions. Concretely, we identify the participants in Lightning as \emph{nodes}. The channels then correspond to edges in a graph. Two nodes open a payment channel through a transaction by individually locking a certain amount of coins on the blockchain to be used for future transactions on the channel. The total capacity of the channel is the sum of the coins that each end-point of the channel locks during the opening transaction and the capacity remains fixed throughout the lifetime of the channel. The initial balance of each node is the amount that the node locked for the opening transaction. However, the balance of each node in the channel changes with payments.

In case of a \emph{dispute} about the current channel balance or a failure to fulfill a previously made commitment, both parties post what they consider to be the latest state on the blockchain. The current mechanism to determine the correct latest state follows a \emph{Replace-by-Revocation} approach. When the two parties agree upon new balances, they sign the new state as well as the revocation of the old state. Hence, parties posting a revoked state as the most recent state can be caught and penalized by losing all their coins in the channel~\cite{lightning}.
At the moment, Lightning is also considering a new dispute resolution, named Eltoo, which should require less storage, easier verification, and the possibility to extend payment channels to more than two people~\cite{decker2018eltoo}. Our attack is independent of the dispute protocol, meaning it will remain applicable even after the change to Eltoo.

\subsection{Lightning Multi-Hop Payments}\label{routing}
\subsubsection{\textbf{HTLC and Payment Execution}}
In a multi-hop payment, the sender sends a payment to a recipient via a path of channels. The sender chooses the path based on its local knowledge of the network topology using one of the three path-finding protocols \emph{LND}, \emph{c-Lightning}, or \emph{Eclair} as we will discuss in detail in Section~\ref{route}. 
Each party along the path commits to make a payment to its successor, which consists of the payment value and all the fees for the succeeding nodes on the path. 
Lightning enforces the commitment of each party to execute the payment through the protocol \emph{hash time-locked contract (HTLC)}~\cite{htlcs}. 

In a HTLC between only two nodes rather than a path, the receiving party of a payment first chooses a random number. It then sends the hash of the random value to the sending party. The two parties then form a contract that if the sending party receives the preimage of the hash within time \emph{timelock}, the sending party will pay the agreed-upon value. The sending party locks the agreed-upon value, i.e., it does not use it for any other payments. If it does not receive the preimage within time \emph{timelock}, the sending party gets the locked amount back.  Disagreements on whether a payment should be made lead to disputes on the blockchain. 
Each channel sets its timelock during its funding transaction and the timelocks are publicly known. 

In a multi-hop transaction, a separate HTLC is set up on each channel of the path.  The recipient of the payment chooses the random value and the hash is distributed to all parties on the path, so that they can set up the HTLCs. 
Note that knowing the hash is sufficient to set up a HTLC. Once all the contracts are set up, the recipient gives the preimage to its predecessor on the path who then forwards it along the path towards the sender to resolve the HTLCs.  The choice of timelocks for the contracts is critical as each party requires sufficient time to receive the preimage from their successor and forward it to their predecessor. 

Concretely, the sender is the first to set up a HTLC with its successor $v$ on the path.  They choose the timelock to be the sum of the timelocks of all the channels on the path.  We call this value the \emph{total timelock} of the channel between the sender and $v$ with regard to the specific HTLC.  Each channel along the path computes its 
\emph{total timelock} by subtracting the timelock of its preceding channel from the total timelock of the preceding channel. 

Now,  a node can only claim the funds from its predecessor if it has the preimage from its successor and hence paid said successor. After receiving the preimage, the time to forward the preimage is exactly the timelock value of preceding channel, giving the intermediary sufficient time to claim its own funds.  
 Thus, Lightning ensures that no honest party loses funds. For our privacy evaluation, it is important to note that all payments use the same hash value, which allows two nodes on the path to determine that they are part of the same payment.

Any intermediary in a payment path can use the total timelock for its outgoing channel to estimate its position with respect to the recipient of the payment, as we will see in Section~\ref{phase1}. As a counter-measure, the BOLT specification of Lightning already has the option \emph{shadow routing}\footnote{\url{https://github.com/lightningnetwork/lnd/issues/1222}}, where the source adds an additional value to its total timelock. The additional timelock value is computed by initiating a random walk from the sender and adding up the timelocks of each channel on the walk. Adding the extra timelock to the last hop creates a large degree of randomness that 
hampers the adversary's ability to predict the recipient of a transaction. 

However, shadow routing is undesired for security reasons: Griefing attacks~\cite{rohrer2019discharged} are severe denial-of-service attacks in which an attacker intentionally causes payment failures at the last possible moment. In this manner, all parties along the payment path have to reserve funds for the maximal amount of time, potentially blocking concurrent payments. The considerable higher timelocks in shadow routing facilitate longer fund reservation and hence increase the severity of the attack. Furthermore, timing attacks still allow inferring the hop distance to the receiver in practice~\cite{nisslmueller2020toward}.

When initiating the payment after determining the total timelock, the sender creates the actual payment message using an onion or layered encryption. The encryption uses the Sphinx package format~\cite{danezis2009sphinx}. The format hides the length of the path as well as the identity of any node on the path excluding the predecessor and successor.  Intermediaries remove a layer of encryption to obtain the identity of the next node on the path.

\subsubsection{\textbf{Route Selection}}\label{route}

The most important aspect exploited by our attack is the selection of the route by the sender. The description of the three route-selection protocols is based upon an extensive code analysis of the Lightning clients $LND$\footnote{\url{https://github.com/lightningnetwork/lnd/blob/31de32686ea3b822dca8c8b84c6f5f3540298ff0/routing/pathfind.go}}, $c-Lightning$\footnote{\url{https://github.com/ElementsProject/lightning}} and $Eclair$\footnote{\url{https://github.com/ACINQ/eclair}}. The crux of the three implementations is that the sender determines the best path using Dijkstra starting from recipient based on a cost function depending on the capacity, charged fees, the locktime, and past experiences with the channel and specified by each implementation for choosing a channel to be added to the path. The reason for starting from the recipient lies in the fact that the charged fees depend on the forwarded value, which includes the fees for the subsequent nodes on the paths. 

Before discussing the route selection algorithms in detail, it is necessary to specify the information the sender has available during route selection. All channels are broadcast upon construction, hence the sender has a topology snapshot including the initial balances of each channel. As the balances change over time, the sender only knows the total to and fro capacity $cap[u,v]$ of the channel $[u,v]$ as well as its age $age[u,v]$. In addition,  nodes’ public keys are distributed. 
For the source to determine the path, it furthermore requires knowledge of the values of fee and timelock $tl[u,v]$.  In Lightning, an intermediary charges two types of fees, a constant base fee $bf[u,v]$ for using the channel and a proportional fee, that is the product of the fee rate $fr[u,v]$ parameter and the amount $amt[u,v]$ to be transferred, where $u$ and $v$ are the channel end-points.
The fee charged by an intermediary $u$ is
\begin{equation}\label{fee}
    fee[u,v] = bf[u,v] + fr[u,v]\cdot amt[u,v].
\end{equation}
The fee parameters can be different for each direction of the channel and any changes to these values need to be broadcast via gossip across the network.


We now discuss the cost functions of the three implementations:

\textbf{LND}: In addition to the fee and timelock parameters of a channel, $LND$'s cost function also has a bias against channels with known recent failures.
    \begin{equation}\label{eq5}
        cost[u,v] = amt[u,v]*tl[u,v]*rf + fee[u,v] +bias[u,v]
    \end{equation}
    Here $rf$ is a risk factor set to $15\cdot 10^{-9}$ by default and $bias[u,v]$ accounts for previous payment failures caused by the channel $[u,v]$. The value of $bias[u,v]$ is extremely large during the first hour after failure and decreases exponentially with every hour elapsed after the last failure.
    If the route fails, the sender is able to track at which channel the failure occurred and marks the time of failure for future reference and calculation of $bias$. A new path is tried until all paths between the sender and recipient have been attempted.
    
\textbf{c-Lightning}: $c-Lightning$ introduces some randomness in the path selection via a parameter $fuzz$ (set to 0.05 by default) and then computes a scaling factor, $scale = 1 + random(-fuzz,fuzz)$.
 Randomization is based on the computation of a one-way 64-bit siphash.
    \begin{equation}\label{eq1}
        cost[u,v] = (amt[u,v]+scale*fee[u,v])*tl[u,v]*rf + bias
    \end{equation}
    where $rf$ and $bias$ are set to 10 and 1 initially. 
    If the path resulting from the first search exceeds the maximal permissible path length of 20, then $rf$ is set to a value close to 0, which essentially turns the search into a shortest path algorithm. A binary search with parameter $bias$ is executed that aims to find a $bias$ that entails a route of permissible length. As the diameter of Lightning is much lower than the maximal permissible length, it is highly unlikely that the adjustments of $riskFactor$ and $bias$ will ever be executed in practice. 
Failed attempts do not entail automatic retries. 

\textbf{Eclair}: 
    $Eclair$ first normalizes the timelock, capacity, and the age by expressing them in relation to their minimal and maximal values. Concretely, the algorithm computes $n_{tl}, n_{cap},n_{age}$ as the corresponding normalized values\footnote{The normalized value $n_D(v)$ of a real number $v$ within range D is computed as $(v-\min{D})/(\max{D}-\min{D})$} between the maximum and minimum admissible values for $tl$ ,$cap$ and $age$,  respectively.
    \begin{align}\label{eq}
    \begin{split}
        cost[u,v] &= fee[u,v]\cdot (n_{tl}[u,v]\cdot tl_{r} \\ 
        &+ (1-n_{cap}[u,v])\cdot cap_{r}  + n_{age}[u,v]\cdot age_{r}). 
    \end{split}    
    \end{align}
    where $tl_{r}=0.15$, $cap_{r}=0.5$, and $age_{r}=0.35$.
    Randomness is introduced by not necessarily choosing the path of the lowest cost. 
    Rather, it chooses among the k-cheapest paths based on Yen's $k$ shortest paths algorithm ~\cite{yen1970algorithm}. The default value of $k$ is set to be 3. Additionally, every payment is retried for default of 5 times until it succeeds.
    
In each of the above three protocols, the sender ensures that the total capacity of each channel in the path is sufficient for the payment. For its own channels, it ensures that the balance is sufficient but it does not know the balance distribution of channels that it is not a part of. Thus, the payment may fail if one of the other channels has an insufficient balance but a sufficient capacity. Nodes on the path prior to the failure have to wait until their respective total timelocks expire to use their committed collateral for other payments.

\section{Adversary Model}
In this section, we discuss the goals and capabilities of our attacker. Before, we introduce our key anonymity notion, the anonymity set, and argue why the anonymity set size is a sufficient metric. 

\subsection{Assumptions}
As described in Section~\ref{sec:background}, Lightning publishes fees, timelocks, and age of channels and nodes, and the clients used by each node. Consequently, we assume such information to be public and uptodate. 

Our main assumption is that the attacker does not have any a-priori knowledge about likely sender-recipient pairs. In other words, all sender-recipient pairs are equally likely. 

We furthermore assume that the attacker does not account for re-routing. Given that re-routing is only possible after timeouts of at least 40 Bitcoin blocks or more than 6h\footnote{\url{https://github.com/lightningnetwork/lightning- rfc/blob/master/ 07- routing- gossip.md}}, it seems likely that people resort to other means to pay. 

\subsection{Anonymity Metrics}

The anonymity set is the set of parties who might cause or have caused an action~\cite{pfitzmann2001anonymity}. In our context, the attacker determines the set of possible senders, recipients, or sender-recipient pairs of a payment. If the size of the anonymity set is small, the anonymity is low. However, the converse is not necessarily true~\cite{serjantov2002towards}. If the probability of parties to be the actual party causing the action varies greatly between parties, the anonymity does not give us sufficient information to claim that a system achieves good anonymity. 

However, for our scenario, there is no reason for such an uneven distribution. With the assumption that all sender-recipient pairs are equally likely,  each sender-recipient pair with a unique cheapest path that fits the observations of the adversary is equally likely to be the one conducting the payment, at least in the absence of multiple cheapest paths. 
If there are multiple cheapest paths, the sender will choose one of them. If only $k$ of $c$ cheapest paths fit the observations of the attacker, the probability for that sender-recipient pair is $k/c$ times the probability of pair with only one cheapest path. However, it seems unlikely that there are a high number of cheapest paths, thus the differences in probability between sender-recipient pairs should be small and hence the anonymity set is still a good metric.

\subsection{Adversary Model}
Given a transaction, the adversary $\att$ aims to identify the sender and recipient. 
Identifying both sender and recipient reveals business relations. Identifying only the sender or receiver still reveals information about buying and selling habits to $\att$. 
While $\att$ might not know the exact transaction value as it only sees the transaction with fees included, these fees are typically low so that the order of magnitude of the transaction value is indeed revealed.

In concrete terms, the attack proceeds as follows: 
Given some transaction information, $\att$  returns two sets: 
\begin{enumerate} 
\item  the set of potential senders $\textbf{S} \subset V$ such that $u \notin \textbf{S}$ is guaranteed not to be the sender of the transaction , and 
\item the set of potential recipients $\textbf{R} \subset V$ such that $v \notin \textbf{R}$ is guaranteed not to be the sender of the transaction. 
\end{enumerate}
$\att$ aims to minimize the size of these sets on average for all possible transactions that goes through $\att$.  

The information $\att$ has on a transaction depends on $\att$'s knowledge and capabilities. 
To emphasize the severity of the vulnerability, we choose a relatively weak adversary. 
The attacker can choose a low number of nodes in the network to compromise.
Allowing the adversary to freely choose the nodes makes sense as $\att$ can likely establish channels with arbitrary nodes if it pays the corresponding fees, an assumption utilized in other state-of-the-art work~\cite{avarikioti2019ride,ersoy2020profit}.
$\att$ is a local internal attacker that can only utilize the information the compromised node is aware of. 
Furthermore, $\att$ acts an honest-but-curious attacker, i.e., they forward transactions and other information as intended by the protocol but aim to derive the above sets from the information. 
$\att$ is non-adaptive and static, i.e., they do not change their attack strategy and do not include past information in their attack. 
Last, $\att$ is a polynomial-time adversary, meaning they are unable to break the applied onion encryption.

\section{Attack Design}\label{attack}
In this section we discuss our attack design. An adversary begins an attack when it observes a transaction as an intermediary. There are two phases in the attack. The first phase involves finding nodes that the adversary can reach with a simple loop-less path that has the same total timelock as observed by the adversary in the actual transaction. The second phase involves curating the list of nodes found in the first phase to compile a list of potential recipients and subsequently a list of potential senders for each potential recipient.

\subsection{Phase I}\label{phase1}
Let us assume that $\textbf{A}$ is an adversary that observes a transaction $T$ as an intermediary. Let $S$ and $R$ be the sender and recipient respectively of $T$, $PRE$ and $NEXT$ be the nodes preceding and succeeding $\textbf{A}$ during the execution of $T$ over $path[T]$, $amt$ be the amount received by $\textbf{A}$ from $PRE$ and $\textbf{TTL}$ be the total timelock from $A$ to $R$. Thus we have,

\begin{center}
    $TTL_{NEXT} = \textbf{TTL} - tl[\textbf{A},NEXT]$ 
\end{center}
and we also have from Equation~\ref{fee}
\begin{center}
    $amt_{NEXT} = \frac{amt - bf[\textbf{A},NEXT]}{1+fr[\textbf{A},NEXT]}$ 
\end{center}
as the total timelock from $NEXT$ to $R$ and the amount reaching $NEXT$, respectively.

We now start a search for all possible loop-less paths starting with $NEXT$, such that the total timelock from $NEXT$ to the recipient of each path is $TTL_{NEXT}$ and the total capacity of each channel in each path is sufficient to forward the transaction. Additionally, we exclude $PRE$ and $\textbf{A}$ from the search. The search is conducted by looking at nodes at all one-hop paths starting at $NEXT$, then all two-hop paths and so on. Paths whose summed timelock of channels exceeds $TTL_{NEXT}$ are excluded. Note that the same node can be visited multiple times during the search via different paths from $NEXT$. Let $P=[P_1,P_2,\ldots,P_n]$ be the list of paths found in the search and $rec_i$ be the final node in $P_i$ and $amt_i$ be the amount reaching $rec_i$ over $P_i$ after accounting for the fees of all nodes preceding $rec_i$. 

If shadow routing is applied, we modify this phase by looking for all loop-less paths without the total timelock restriction. This would essentially consider all paths from $NEXT$ to every node in the network other than $\textbf{A}$ and $PRE$. 

\subsection{Phase II}
Let $P_i \in P$ be a path found in \textit{Phase I} and $P_i = \{p_1,p_2,\ldots,p_r\}$ with $p_1 = {NEXT}$ and $p_r = rec_i$.  We append $PRE$ and $\textbf{A}$ to each $P_i$ to get $P'_i = \{PRE,\textbf{A},p_1,p_2,\ldots,p_r\}$. We now determine if $rec_i$ is indeed a potential recipient of $T$ and if so, then we determine all potential senders that could have made a payment to $rec_i$. 

    \textbf{Step 1}: The algorithm computes  paths from all nodes in the network to $rec_i$ using the cost functions of  $LND$, $c$-$Lightning$ or $Eclair$. We find the 3 best paths from all nodes to $rec_i$ in case of $Eclair$.  The paths are computed such that the first node $N$ of the computed part is treated as an intermediary, i.e., it charges a fee.  In other words, we are determining potential second nodes one the path. Focusing on the second nodes first makes sense as the decision on the first channel includes information about the channel balance, which is unknown to the attacker. 
    
    \textbf{Step 2}: Let $P[N]$ be the path computed from $N$ to $rec_i$. Let $p_j$ be a node in $P_i$ and $P_i[p_j:]$ be the sub-path of $P_i$ excluding the nodes from $p_1$ to $p_{j-1}$. If $P[p_j]$ is not equal to $P_i[p_j:]$, it means that the path $P'_i$ would not have been computed during a path computation and hence $rec_i$ cannot be a destination. Once we establish that the sub-path of $P'_i[\textbf{A}:]$ is the same as $P[\textbf{A}]$, $rec_i$ can be considered as a possible recipient of $T$ and we can proceed to find potential senders. For $Eclair$, we dismiss $rec_i$ as a potential recipient if none of the 3 best paths match the sub-path of $P_i$ from $p_j$ to $p_r$. 
    
    \textbf{Step 3}: If $P[pre]$ is not equal to $P'_i$, then $PRE$ cannot be an intermediary. However since we know that $PRE$ did precede $\textbf{A}$, $PRE$ has to be the sender of the transaction. We can then say that $PRE$ is the only possible sender if $rec_i$ is the recipient. If $P[PRE]$ is indeed equal to $P'_i$, then we consider $PRE$ as just one of the potential senders if $rec_i$ is the recipient and continue. In the case of $Eclair$, we claim that $PRE$ is the only possible source if all 3 paths do not match $P'_i$. If any one of them does, we add $PRE$ as one potential sender.
    
    \textbf{Step 4}: If for any node $N$, we find that $P[N]$ has $P'_i$ as a sub-path, then we add all neighbors of $N$ that are not in $P[N]$ as a potential senders as we can now confidently claim that $N$ is a potential second node in $path[T]$. For $Eclair$, if any one of the best 3 paths has $P'_i$ as a sub-path, then we add $N$'s neighbors as potential senders. 
For enhanced performance of Eclair, we did not compute all three paths if a match was found for the first or second path in Step 2--4. 

Execute the above steps for all $P_i$ in $P$ for a list of potential recipients as the recipient anonymity set $\textbf{R}$ and potential senders for each such recipient. The union of the potential senders for all potential recipients is the sender anonymity set $\textbf{S}$. 

\subsection{Colluding Adversaries}
There are ways to extend the attack to multiple adversaries. One way is to have multiple adversaries independently execute the attack. However, attackers can also combine their observations.
Concretely, two adversarial nodes determine if they observe the same transaction based on the hash value used in the HTLC, as pointed out in Section~\ref{sec:background}.  
After each adversary complete their respective anonymity sets, we take the intersection of all destination anonymity sets as the final recipient set. Analogously, we derive the intersection of the sender anonymity sets. 

We note that recent work by Malavolta et al. ~\cite{malavoltamulti} would remove this straight-forward linkability if it was implemented in Lightning. Our results without collusion remain applicable though.

\subsection{Attack Complexity}
Phase I of the attack is essentially a problem of finding all the simple paths between $NEXT$ and every node in the network other than $PRE$ and $\textbf{A}$ with a complexity of $O(|V|!)$ in the worst case~\footnote{https://www.baeldung.com/cs/simple-paths-between-two-vertices} and cannot be run in polynomial time. To avoid this excessive run-time, we optimize the attack to search for paths only up to a hop-count of $d$ from $NEXT$, resulting in a worst case complexity of $O(|V|^d)$ as we could end up having every node in the network at each hop in the worst case with a complete graph. Since the graph of Lightning is not a complete graph, a more realistic complexity is $O((Deg_{max})^d)$ where $Deg_{max}$ is the maximum degree of any node in the network.

If the search is incomplete at hop-count $d$, then recipient anonymity set could be incomplete as well and there is a chance that we miss the actual recipient if the recipient was indeed at a hop-count greater than $d$ from $NEXT$. In the case of colluding adversaries, it maybe the case that one or more adversaries fail to include the actual recipient in their anonymity sets. To account for this case, we also consider the intersection of the anonymity sets obtained by only those adversaries whose search in Phase I reached a natural conclusion before depth $d$.


During Phase II, we compute cheapest paths from all nodes to each $rec_i$ and is the same as running single-source Dijkstra with priority queue with a run-time of $\mathcal{O}((V+E)log(V))$,  when using a priority queue using binary heaps. Thus, the total run-time for all $rec_i$ is $\mathcal{O}(V(V+E)log(V))$. The total complexity of the attack is the product of the complexities of Phase I and Phase II, that is $\mathcal{O}((Deg_{max})^d)(V+E)V log(V))$.


\section{Evaluation}
In this section, we leverage the described attack to determine
anonymity sets.  

\subsection{Metrics}\label{metrics}
To evaluate our attack we use the following metrics:
\begin{itemize}
	\item Size of the sender and recipient anonymity sets.
	\item The proportion of transactions attacked $R_{att}$,  and the average number of attacks per attacked transaction $Av_{att}$ to indicate the number of transactions attacked by two or more adversaries. 
	\item The correlation between the size of the sender anonymity set and the distances $CorrD_{S}$ and $CorrD_{R}$ between the adversary and the sender and receiver, respectively. 
	\item The proportion of attacks that have a singular sender anonymity set, $Sing_S$, a singular recipient anonymity set, $Sing_R$, both a singular sender and recipient anonymity set, $Sing_{both}$, and at least one singular anonymity set,$Sing_{any}$. 
	\item The proportion of attacks $NatEnd$ for which Phase I of the attack reached its natural conclusion before depth $d$.
	\item The proportion of transactions $Comp_{att}$ for which the correct recipient was included in the anonymity set.
\end{itemize}

Note that the correct recipient is always included if the attack was not aborted. Otherwise, it might not have been included if it was at a distance greater than $d$ from $NEXT$.   

\begin{figure*}
    \centering
    \vspace{-2.5em}
    \subfloat[\emph{\textbf{Sender anonymity - LND only}}]{\includegraphics[scale=0.12]{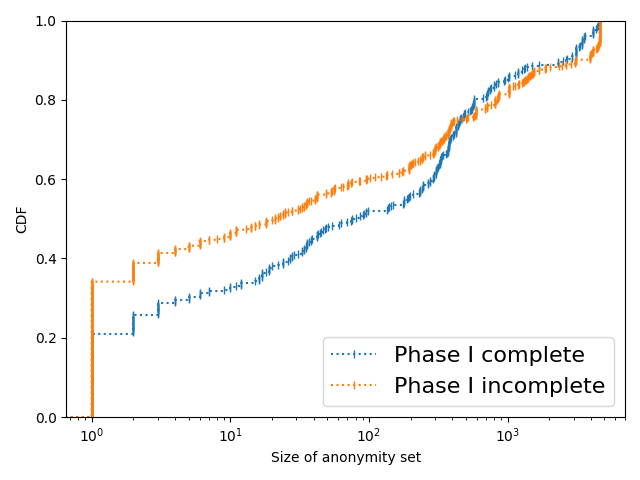}\label{source_lnd}} \hfill 
    \subfloat[\emph{\textbf{Sender anonymity - clients known}}]{\includegraphics[scale=0.12]{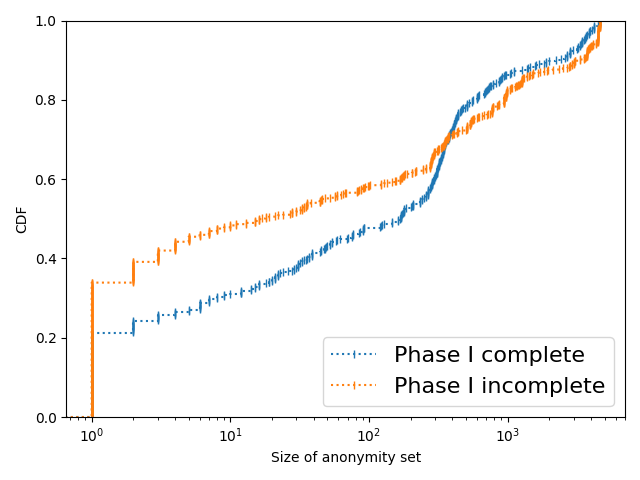}\label{source_mixed} }\hfill
    \subfloat[\emph{\textbf{Sender anonymity - blind}}]{\includegraphics[scale=0.12]{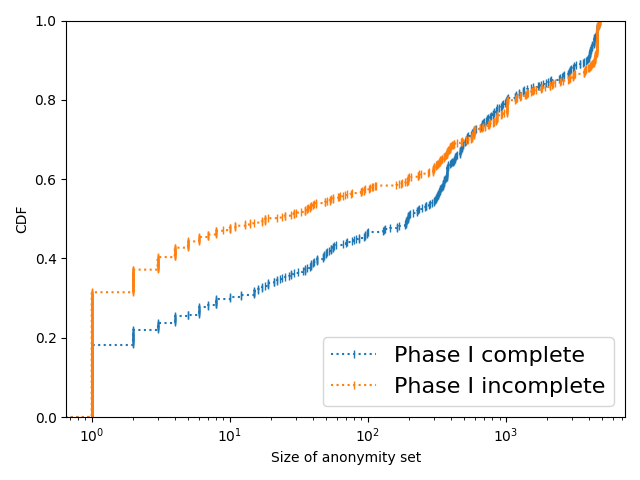}\label{source_blind} }\hfill
    \subfloat[\emph{\textbf{Recipient anonymity - LND only}}]{\includegraphics[scale=0.12]{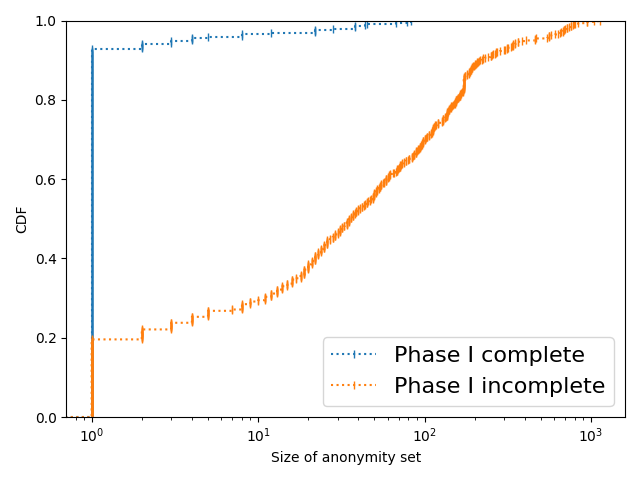}\label{dest_lnd} }\hfill
    \subfloat[\emph{\textbf{Recipient anonymity - clients known}}]{\includegraphics[scale=0.12]{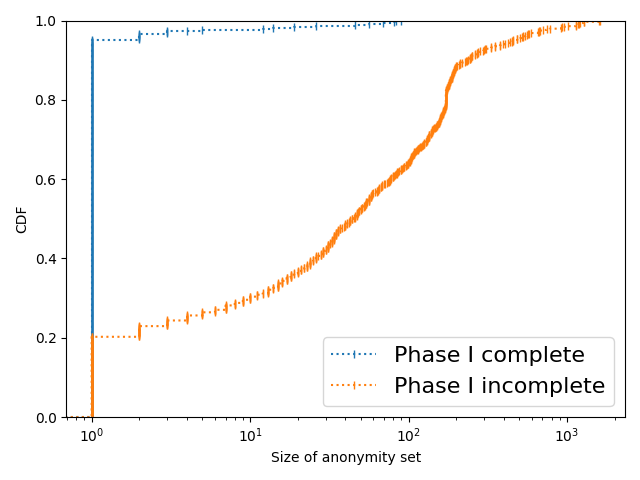}\label{dest_mixed} }\hfill
    \subfloat[\emph{\textbf{Recipient anonymity - blind}}]{\includegraphics[scale=0.12]{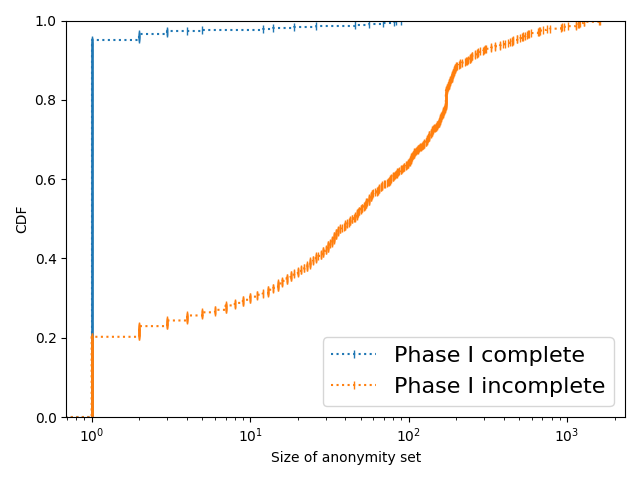}\label{dest_blind} }\hfill
    \vspace{-1em}
    \caption{Distribution of the sizes of the sender and recipient anonymity set sizes.}
    \label{fig:1}
    \vspace{-2em}
\end{figure*}

\subsection{Data-sets and Parameters}\label{data}
We use a snapshot of the Lightning Network obtained in June 2020 from \url{https://ln.bigsun.xyz} to evaluate our attack.  The snapshot had a total of 11197 nodes and 82989 channels. We do not include inactive nodes and channels that have been closed. After eliminating closed channels and inactive nodes, we ended up with a network having 4791 nodes and 28997 channels. We randomly distributed the capacity between the endpoints of each channel. 51\% of the timelocks are 144 blocks, 25\%
are 40 blocks and 10\% are 30 blocks. 
All channels publicly revealed the clients that they are using for routing payments. Around 92\% of these nodes use lnd, 6\% use
c-lightning and 2\% use eclair. 

We evaluate our attack in three settings:
\begin{itemize}
	\item \textbf{LND only}: All nodes only use $LND$ as their client for routing. This is a good baseline as nearly all nodes use $LND$ as their client.
	\item \textbf{clients known}: All nodes use their assigned client as per the snapshot and the adversary know the client that each node is using as is seen in reality.
	\item \textbf{blind}: The three clients are distributed between the nodes with the same distribution as seen in the snapshot. However, the adversary is not aware of the clients used by any node. We evaluate this scenario to account for a future scenario where nodes do not publish their routing clients.
\end{itemize}

In each of the settings, the sender and recipient for each transaction are assigned randomly and the transaction amount is distributed exponentially between 1 and 100000 satoshis and 1000 such transactions are simulated. The depth $d$ is set to be 3 for the attack so that recipients at a distance of upto 4 hops from the adversary are included in the recipient anonymity set.

\textbf{Choosing adversaries}: Lightning has been observed to be a small-world and scale-free graph~\cite{rohrer2019discharged}. Recent work furthermore indicates an increasing trend to centralization, meaning that most paths pass the same small set of central nodes~\cite{hong2020centralization}. 

Hence, the position of adversarial nodes has critical impact on the proportion of transactions attacked and the size of the anonymity sets. Central adversarial nodes observe many payments and hence there is a high chance that the payment passes through them and they have a chance to launch the attack. However, as a large number of sender-recipient pairs utilize the node, the anonymity sets might be quite large. 
In contrast, adversarial nodes that have a low centrality will observe only a very small fraction of payments. Yet, as the set of senders who choose this node is so small, the anonymity set sizes are also expected to be smaller. Introducing a central adversary would also entail higher monetary collateral as they have more channels. We found that the node with the highest capacity has a total capacity of around 128 BTC and top 10 nodes with the highest capacities in Lightning have an average capacity of around 54 BTC\footnote{https://1ml.com/node?order=capacity}. So, we assume that the cost of introducing a well-connected adversary to be around 54 BTC.  In contrast, on average, a node only has a capacity of 0.115 BTC, so introducing a node without a high centrality is comparably cheap. 

As a consequence, we experiment with choosing adversarial nodes randomly, by highest centrality, or by lowest centrality. 
Centrality metrics considered in our evaluation are \emph{betweenness}, \emph{closeness}, \emph{degree} and \emph{eigen vector} centrality metrics.
Note that for the betweenness centrality, we consider only the the shortest paths without weights. A total of 21 nodes were assigned to be adversaries. We found that the top 6 nodes for each centrality metric to be the same and we used them as adversaries. To ensure diversity, 5 nodes with low centralities and 10 random nodes were also used as adversaries. Note that nodes with the lowest centralities were those that had only one connection and hence were not used. Instead, we choose 5 nodes from a list of 20 nodes with lowest centralities (5 for each metric) among the nodes having more than one channels.

\subsection{Simulation Model}
To simulate our attack, we implemented the routing algorithms of
all three Lightning clients \emph{LND}, \emph{c-Lightning}, and \emph{Eclair}.  In the case of \emph{Eclair},  we use a generalized version of Dijkstra to compute the 3 best paths instead of using Yen's algorithm for simplicity as Yen's algorithm is slower than Dijkstra when we are not dealing with negative weights. We do not
include the cryptographic aspects since we are not attacking the
the cryptographic contracts between channels and the blockchain
here. 

For each transaction, only a single attempt is made for routing given that the total
timelock is most likely in excess of 40 blocks ( 6 hours) and nodes are
unlikely to continue trying after such a delay. Thus, we disregard
the notion of edge probability bias for past failures that is used in cost function of \emph{LND}.  On receiving a transaction to forward, the adversary first forwards the payment if its balance permits and then  computes the anonymity sets for the sender and the recipient of the transaction.  The adversary computes the anonymity sets only on receiving a transaction. If the transaction is delayed or a timeout failure occurs before reaching the adversary, then the adversary simply does not compute the anonymity sets. If the transaction is delayed or a timeout failure occurs after reaching the adversary, the adversary still has the necessary information to compute the anonymity sets. 
We execute transactions sequentially as concurrency does not affect the route selection or attack as such, only the probability that a transaction fails before reaching the adversary is increased. 

All simulations were executed using \textit{Python 3}. We used the \textit{networkx} module to handle graph related operations. The source code can be found on GitHub~\footnote{\url{https://github.com/SatwikPrabhu/Attacking-Lightning-s-anonymity}} .

\subsection{Results}
We simulate the attack for the three settings discussed in Section~\ref{data} and plot the cumulative distribution function of the size of the sender and recipient anonymity sets.

\begin{figure*}
    \centering
    \vspace{-2em}
    \subfloat[\emph{\textbf{Sender anonymity - LND only}}]{\includegraphics[scale=0.12]{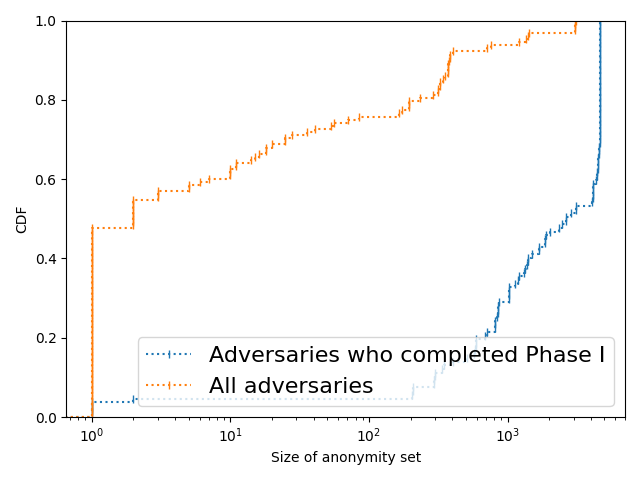}\label{coll_source_lnd}} \hfill 
    \subfloat[\emph{\textbf{Sender anonymity - clients known}}]{\includegraphics[scale=0.12]{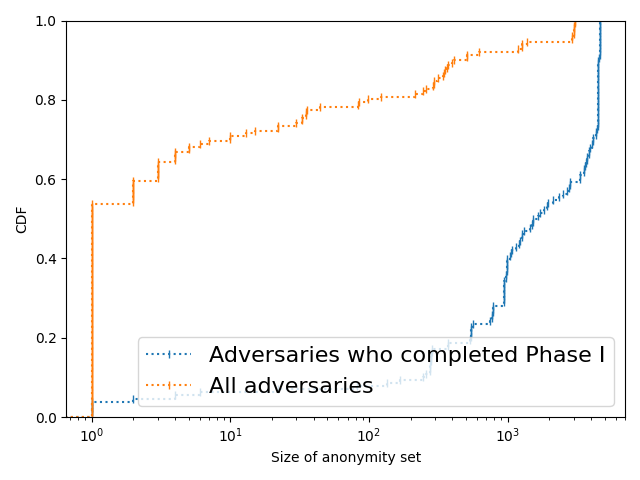}\label{coll_source_mixed} }\hfill
    \subfloat[\emph{\textbf{Sender anonymity - blind}}]{\includegraphics[scale=0.12]{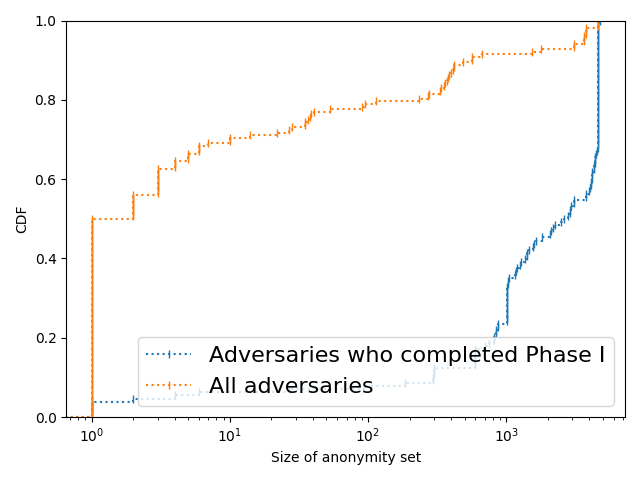}\label{coll_source_blind} }\hfill
    \subfloat[\emph{\textbf{Recipient anonymity - LND only}}]{\includegraphics[scale=0.12]{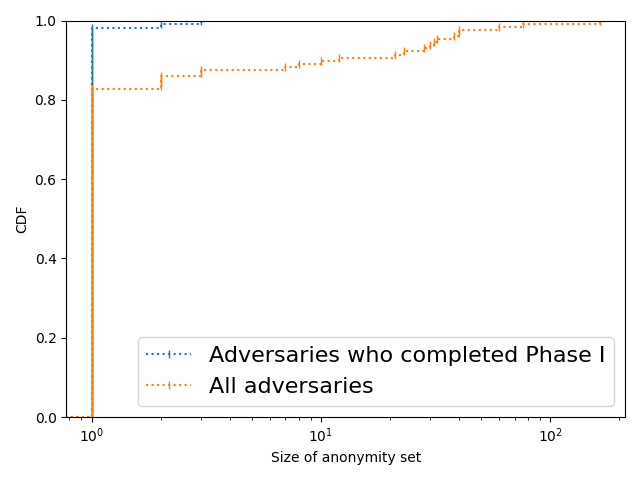}\label{coll_dest_lnd} }\hfill
    \subfloat[\emph{\textbf{Recipient anonymity - clients known}}]{\includegraphics[scale=0.12]{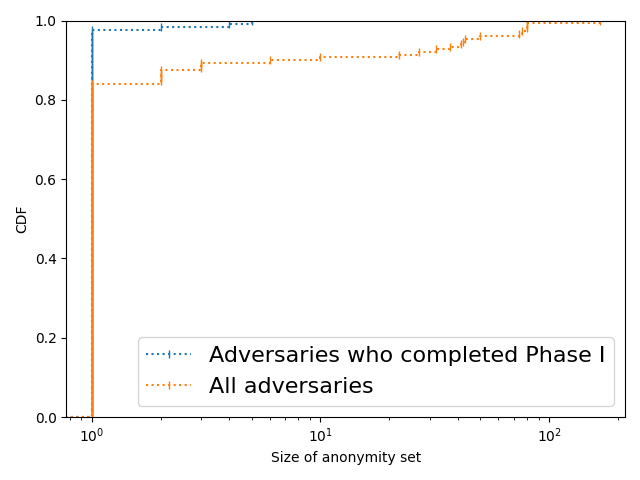}\label{coll_dest_mixed} }\hfill
    \subfloat[\emph{\textbf{Recipient anonymity - blind}}]{\includegraphics[scale=0.12]{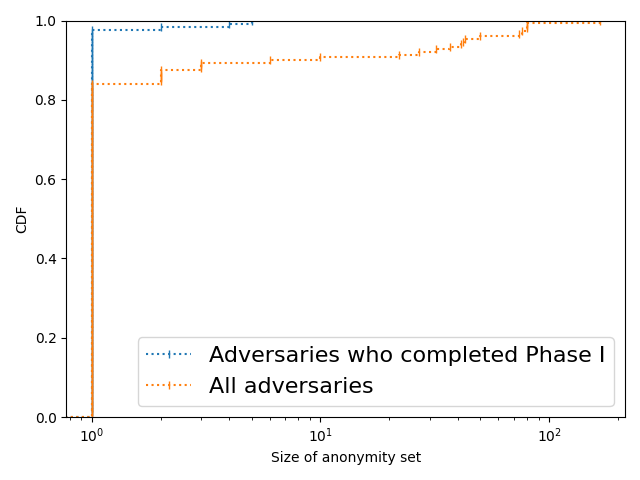}\label{coll_dest_blind} }\hfill
    \vspace{-1em}
    \caption{Distribution of the sizes of the sender and recipient anonymity set sizes only for colluded attacks.}
    \label{fig:2}
    \vspace{-2em}
\end{figure*}

Figure~\ref{fig:1} shows the sizes of the sender and recipient anonymity sets as a cumulative distribution function, differentiating between the cases when Phase I of the attack was completed naturally or ended at depth $d$ from $NEXT$.

From Figure~\ref{dest_lnd}, \ref{dest_mixed} and~\ref{dest_blind}, we can see that the recipient anonymity set is singular in excess of 90\% cases when Phase I of the attack is completely executed and in around 20\% cases when Phase I of the attack is incomplete for all three settings. Moreover, the size of the recipient anonymity set is less than 100 in around 60\% cases regardless of whether Phase I was completed or not. In general, the size of the recipient anonymity set is lower in the cases where Phase I was completed. This is because $TTL_{NEXT}$ is lower when Phase I is completed and only nodes close to the adversary need to be considered.

From Figure \ref{source_lnd}, \ref{source_mixed} and \ref{source_blind}, we can see that the size of the sender anonymity set is singular in nearly 40\% cases when Phase I is incomplete and in around 20\% cases when Phase I is completed for all three settings. This is because Phase I might not be completed when the adversary is closer to the sender than it is to the recipient, leading to the size of the sender anonymity set being generally lower than when Phase I was completed.

\begin{table}[ht]\label{table}
\centering
\caption{Attack metrics}
\vspace{-1em}
\begin{tabular}{ |c|c|c|c| }
    \hline
    Test case&\textbf{LND only}&\textbf{clients known}&\textbf{blind}\\
    \hline
    $R_{att}$&0.54&0.51&0.51\\
    \hline
    $Av_{att}$&1.19&1.19&1.19\\
    \hline
    $CorrD_S$&0.78&0.73&0.79\\
    \hline
    $CorrD_R$&0.46&0.47&0.47\\
    \hline
    $Sing_S$&0.29&0.28&0.26\\
    \hline
    $Sing_R$&0.49&0.50&0.50\\
    \hline
    $Sing_{any}$&0.67&0.70&0.69\\
    \hline
    $Sing_{all}$&0.08&0.08&0.07\\
    \hline
    $NatEnd$&0.47&0.51&0.51\\
    \hline
    $Comp_{att}$&0.99&0.99&0.99\\
    \hline
\end{tabular}
\vspace{-1.3em}
\label{table1}
\end{table}

We summarize the remaining metrics from Section~\ref{metrics} in Table~\ref{table1}. The results for all three settings are similar. The key results are as follows:
\begin{itemize}
    \item More than 50\% of the transactions are observed by at least one adversary. 
    \item The size of the sender anonymity set is singular in around 30\% of attacks, the size of the recipient anonymity set is singular in around 50\% of the attacks. One of the two sets is singular in around 70\% of the attacks and both are singular in around 8\% of the attacks.
    \item The adversary is successful in finding the recipient within 3 hops from $NEXT$ in at least 99\% of attacks and consequently finds the sender as well in these cases.
    \item Phase I of the attack reaches its natural conclusion for around 50\% of the attacks.
    \item There is a strong positive correlation between the size of the sender/recipient anonymity sets and the hop-count between the adversary and the sender/recipient.
\end{itemize}


Figure~\ref{fig:2} shows the impact of the colluded attack. For transactions that were attacked by multiple adversaries, we show the sizes of the intersection of the sender anonymity sets returned by each adversary and similarly for the recipient anonymity sets. We also separately show the sizes of the intersections ignoring the anonymity sets returned by adversaries without completing Phase I of the attack. 

From Figure~\ref{coll_source_lnd}, \ref{coll_source_mixed} and \ref{coll_source_blind}, we can see that the size of the sender anonymity set is 1 in around 50\% cases and is less than 10 in excess of 60\% cases when we do not ignore all adversaries who may not have completed Phase I. However, the size of the sender anonymity sets are consistently higher when we ignore the adversaries that have not completed Phase I since the adversaries that do complete Phase I are likely to be very close to the recipient and away from the sender in nearly all cases. From Figure~\ref{coll_dest_lnd}, \ref{coll_dest_mixed} and \ref{coll_dest_blind}, we can see that the size of the recipient anonymity set is consistently 1 when we include all attacker nodes. We also see that the size is generally less than 100 if we consider the sets returned by all adversaries, showing that a colluded attack is very effective in predicting the recipient of a transaction. 

\begin{figure}
    \centering
    \vspace{-0.5em}
    \subfloat[\emph{\textbf{Sender anonymity - LND only}}]{\includegraphics[scale=0.14]{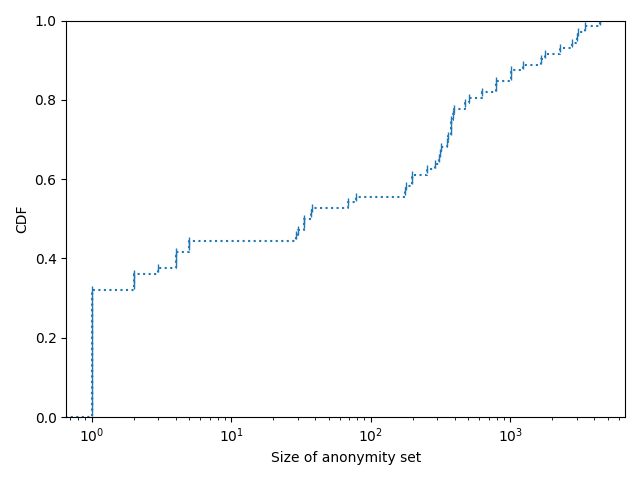}\label{shadow_dest}} \hfill 
    \subfloat[\emph{\textbf{Recipient anonymity - LND only}}]{\includegraphics[scale=0.14]{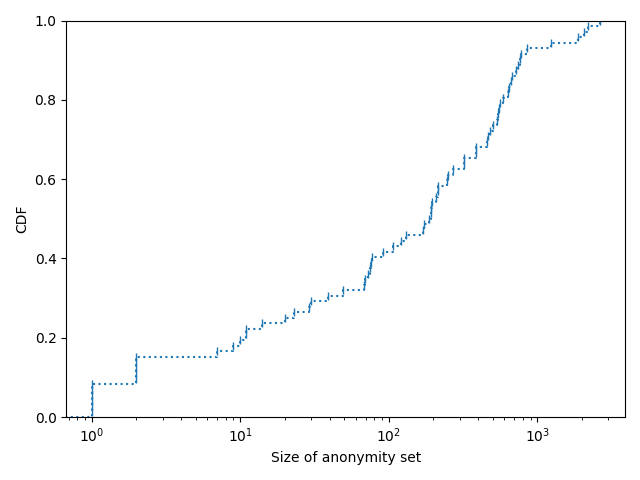}\label{shadow_source} }\hfill
    \vspace{-1em}
    \caption{Distribution of the sizes of the sender and recipient anonymity sets obtained assuming shadow routing is used}
     \label{fig:3}
     \vspace{-2em}
\end{figure}

Additionally, we depict the sizes of the anonymity sets when shadow routing is used in Figure~\ref{fig:3}. However, due to the high computational overhead of computing all paths without having a timelock value as a stopping criteria, we set the depth $d$ to be 2 and only simulate 100 transactions in the \textbf{LND only} setting. From Figure~\ref{shadow_dest}, we see that the recipient anonymity sets are much larger compared to the standard routing algorithm. This is expected as we do not have the timelock restriction to narrow down the set of recipients. In contrast,as shown in Figure~\ref{shadow_source}, the sizes of the sender anonymity sets are similar to the case of standard routing. This similarity is expected because shadow routing does not affect Phase II of the attack when we find potential senders.

Our results show that well-connected nodes in the network under adversarial control are capable of observing a sizeable portion of transactions in the network. Using our de-anonymization attack, a unique sender and/or recipient can be determined quite regularly. Moreover, adversaries can collude in the current version of Lightning to share their respective anonymity sets and reduce the anonymity further. We also show that the attack performs equally well when the adversary doesn't know the client used by the sender. Shadow routing can be used to mitigate the risks of the attack. Although shadow routing is successful is increasing the recipient anonymity, it does not seem to similarly increase the sender anonymity. 

\section{Related Work}
\label{sec:related}

There are two areas of related research for this topic: i) privacy in payment channel networks, with the focus on work for Lightning, and ii) anonymous source routing in other contexts. 

\subsection{Payment channel network anonymity}
Recent work presented three attacks on Lightning anonymity. Tikhomorov et al.\ show that sender and recipient of a transaction can be de-anonymized if the nodes succeeding the sender and preceding the recipient are compromised by an adversary. The two compromised nodes can then determine that they are part of the same transaction using the cryptographic challenge included in the HTLC of the same payment and hence identify the sender and the recipient. The authors evaluate the proportion of paths between any pair of nodes that can be de-anonymized when well-connected nodes are compromised. However, the attack requires at least two compromised nodes on the path in contrast to our attack~\cite{tikhomorov_snp}. 

Nisslmueller et al.\  investigate timing attacks to get the identity of the receiver by measuring the time between locking collateral at an intermediary and the time when receiving the preimage to unlock the collateral~\cite{nisslmueller2020toward}. In contrast to our approach, they were unable to determine the sender and their results are only applicable when jitter is low. The attack could furthermore be prevented by adding random delays at the receiver, at the cost of locking collateral slightly longer. Given the extremely long periods collateral is locked for failed payments, adding delays of less than a second to prevent a timing attack seems like an appropriate solution. 
In general, their attack is complementary to ours and can replace estimating the receiver's position if Shadow Routing is applied. 

The third approach by  Kappos et al.\ leverage the fact that paths in Lightning currently tend to be short, i.e., a lot of paths only have one intermediary. As a consequence, they evaluate the effectiveness of an attacker that always assume its predecessor is the sender and its successor is the recipient~\cite{kappos2020empirical}. In contrast to our attack, the attack hence merely assigns one node a probability to be the sender or recipient rather than determining the complete anonymity set. 
Furthermore, they are unable to be certain about the sender's or recipient's identity while our approach allows in many cases to narrow the possible senders and receivers down to one party. 

Previous work on improving the anonymity of payment channel networks led to the development of anonymous multi-hop locks~\cite{malavoltamulti}. 
When considering colluding attackers, we assumed that the attackers can infer if they are on the same path due to the shared hash value for the HTLC. Anonymous multi-hop locks replace the constant hash value with a randomized and unlinkable value, meaning that the colluding attackers are unable to easily tell if they are on the same path. However, side channels such as transaction values and timing are still available alternatives.
Note that even without multiple adversaries on the path, we decrease the anonymity considerably. 

Multiple works suggest changing to an inherently different routing algorithm to preserve anonymity~\cite{malavolta2017silentwhispers,roos2017settling,malavolta2017concurrency}. However, all of the approaches still choose paths strategically rather than completely randomly. They express their privacy guarantees in terms of shared paths and topology features but it remains unclear how these guarantees relate to concrete anonymity metrics.  Due to the importance of choosing paths that have sufficient capacity and are preferably both short and cheap, any routing algorithm is a trade-off between privacy and these performance measures.
A detailed quantitative evaluation of the proposed algorithms in terms of these trade-offs is necessary before deploying them as a replacement for Lightning's routing. Our attack can likely be adapted for such an evaluation.

In addition to anonymity, 
there exist attacks to infer the balance of channels. 
In contrast to our attacks, these are active attacks that attempt to 
route payments of differing values through channels~\cite{kappos2020empirical,nisslmueller2020toward,tikhomirov2020probing,herrera-joancomarti-balances}.
Our attack could be improved by having knowledge about the channel balances, e.g., we wouldn't need to add all neighbors of a potential first hop as potential senders.
So, while the attacks are fundamentally different both in goal and execution, they can likely help to further increase the strength of our attack.

Further attacks on Lightning and payment channel networks indicate that denial-of-service attacks are easily possible~\cite{rohrer2019discharged,herrera-joancomarti-balances}. In particular, Tochner et al.\ relate the susceptibility of Lightning to denial-of-service attacks to the predictability of routing. They find that a combination of adding noise to the fee and choosing one of the top cheapest paths mitigates denial-of-service resistance~\cite{tochner2019hijacking}.

\subsection{Anonymous Source Routing}
\label{sec:anonymoussource}

Anonymity in source routing has been discussed in the context of network layer anonymity for alternative Internet architectures~\cite{hsiao2012lap,sankey2014dovetail,chen2017phi,chen2015hornet,chen2018taranet}.
These works face the same problem as Lightning's routing in the sense that the topology is restricted, i.e., it is not possible to choose random nodes for the path, and performance is highly important. However, routing packets for communication does not require intermediaries to have information about transaction values, fees, and timelocks. Hence,  these protocols should leak less information and hence naturally achieve bigger anonymity sets.  
Yet, while the protocols are not directly applicable for Lightning, some of the concepts of anonymous source routing protocols could be promising ideas for improving Lightning's anonymity. 
The existing works on anonymous source routing can be grouped into three classes: i) alternative packet formats, ii) timing attacks, and iii) alternative path selections. 

Phi~\cite{chen2017phi} and Hornet~\cite{chen2015hornet} design alternative packet formats to hide explicit information about the path chosen by the source such as the length. Like Lightning, Hornet uses Sphinx~\cite{chen2015hornet} to achieve the same goals.  Hence, Lightning's routing protocol already has the protections proposed by Phi and Hornet. 

Our attack  remains applicable even in the presence of defenses against timing attacks. However,  work from that area might be helpful in overcoming proposed timing attacks on Lightning~\cite{nisslmueller2020toward}. 
Concretely, Taranet suggests re-ordering of traffic and the insertion of dummy traffic~\cite{chen2018taranet}.  

The only class of protocols directly related to our attack are those that propose path selection protocols that are less easily predictable. The two key works in the area of source routing are LAP and Dovetail. 
LAP only achieves anonymity if the first node on the path is not compromised~\cite{hsiao2012lap}, which is not a suitable assumption for Lightning. Nodes are likely to connect to business partners who should not be trusted to know about other transactions of their neighbors. 
Dovetail improves upon LAP through the use of two additional nodes, a matchmaker and a dovetail node. In a nutshell, the sender uses the matchmaker node as the recipient and the recipient has the matchmaker as the sender. The algorithm then trims the path such that parts that are shared between the two paths from sender to matchmaker and matchmaker and receiver are removed. The dovetail node is the only node on the resulting loop-free path that was contained in both paths~\cite{sankey2014dovetail}. 
Integrating Dovetail in Lightning is an avenue for future work, which we will discuss in more detail in Section~\ref{sec:conclusion}.

\section{Conclusion and Future Research}
\label{sec:conclusion}

Our experiments indicate that the level of anonymity in Lightning is low. Merely using the layered encryption of onion routing does not guarantee substantial anonymity in the presence of an almost deterministic path selection. 

While \emph{shadow routing} mitigates the risks of an intermediary determining its position in a transaction by adding extra timeout delay, it comes at an additional risk of payments being locked for a longer time and hence increasing the existing scope for griefing attacks. Hence, the question remains on how to design routing protocols that achieve acceptable performance, fees, security and privacy. 

As discussed in Section~\ref{sec:anonymoussource}, Dovetail~\cite{sankey2014dovetail} is a promising option for increasing anonymity. However, in Dovetail, there are two partial paths and messages on these paths are supposed to be unlinkable. In Lightning, it might be possible to link payments due to metadata such as payment values and timelocks. A thorough evaluation is required to determine the degree of linkability and the impact of Dovetail's longer path on payment success and susceptibility to griefing attacks. 
In the end, a more light-weight protocol could be the best option for Lightning. Simple options like adding a random short detour to a chosen path might not achieve perfect unlinkability but improve the current level sufficiently without creating severe performance or security issues. 
We plan to evaluate both light-weight options and Dovetail in terms of anonymity, performance, and security. 
Furthermore, we aim to apply our methodology to quantify the anonymity of alternative routing protocols proposed for Lightning. 

\begin{acks}
The work was partially supported by Ripple's University Blockchain Research Initiative and the Distributed ASCI Supercomputer (DAS) was used for experiments.
\end{acks}

\bibliographystyle{ACM-Reference-Format}
\bibliography{sample-authordraft}


\end{document}